\documentstyle[aps]{revtex}
\begin{document}

\narrowtext
\twocolumn


\noindent
{\bf Big bang simulation in superfluid  $^3$He-B --- Vortex nucleation in
neutron-irradiated superflow}
\vspace{6mm}\\
\noindent
 V.M.H. Ruutu$^1$, V.B.  Eltsov$^{1,2}$, A.J. Gill$^{3,4}$, T.W.B. Kibble$^3$,
M. Krusius$^1$, Yu.G. Makhlin$^{1,5}$, B. Pla\c{c}ais$^6$, G.E.
Volovik$^{1,5}$, and Wen Xu$^1$\\

\footnotetext{$^1$ Low Temperature Laboratory, Helsinki University of
Technology, 02150 Espoo, Finland.
\\  $^2$ Kapitza Institute for Physical Problems, 117334 Moscow.\\
$^3$ Blackett Laboratory, Imperial College, London SW7 2BZ.\\
$^4$ T-6 Theoretical Division, Los Alamos National Laboratory,
Los Alamos, NM 87545.\\
$^5$ Landau Institute for Theoretical Physics, 117334 Moscow.\\
$^6$ Laboratoire de Physique de la Mati\`{e}re Condens\'{e}e de
l' Ecole
Normale Sup\'{e}rieure, F-75231 Paris CEDEX 05.
}

\vspace{6mm}
\noindent


{\bf We report the observation of vortex formation upon the absorption of a
thermal neutron in
a rotating container of superfluid $^3$He-B. The nuclear reaction n + $^3_2$He
=
p + $^3_1$H + 0.76MeV heats a cigar shaped  region of the superfluid into
the normal phase. The
subsequent cooling of  this region back through the superfluid transition
results in the
nucleation of  quantized vortices.  Depending on the superflow velocity,
sufficiently large
vortex rings grow  under the influence of the Magnus force and escape into
the container
volume where they are detected individually with nuclear
magnetic resonance. The larger the
superflow velocity the
smaller the  rings which can expand. Thus it is possible to obtain
information  about the
morphology of the initial defect network. We suggest that the nucleation of
vortices during
the rapid cool-down into the superfluid phase is similar to the formation
of defects during
cosmological phase transitions in the early universe.}

\vspace{6mm}


An issue of current interest is the generation of topological defects during
symmetry-breaking  phase transitions \cite{Kibble}. In our experiment we
study the density of
vortex lines  produced during a temperature quench of normal $^3$He liquid
into its superfluid
B-phase. A sufficiently rapid transition  results in a dense network of
vortex lines
\cite{Zurek}. In a large enough bias field of superfluid flow some vortices
will be
stabilized and survive the
subsequent relaxation towards equilibrium while others contract and annihilate.
Current theories of condensed matter physics generally provide the tools
for understanding defect
nucleation in slow phase transitions close to equilibrium. The present
nucleation process, however, is far out of equilibrium and resembles the
discussion of
cosmological phase transitions in the early universe \cite{Review2}. Primordial
defect nucleation may explain the cosmological large scale structure and
also in some
scenarios the net baryon asymmetry of the present universe. To test the
validity of such
estimates  it is important to find laboratory experiments in which defect
formation in a rapid quench can be studied.

Liquid $^3$He has several  advantages over other systems which in the past
have served  as laboratory
models for the cosmological defect-formation scenario, such as liquid crystals
\cite{nematic}, liquid $^4$He \cite{helium}, and  superconductors
\cite{ZurekReview}. Many direct
parallels and formal analogies connect superfluid $^3$He theory with field
theories which
are used to describe the physical vacuum, gauge fields or fermionic elementary
particles \cite{Volovik}. Superfluid $^3$He exhibits a variety of phase
transitions and
topological defects which can be detected with NMR methods with
single-defect sensitivity
\cite{Nucleation}. Of particular relevance to our rapid quench experiment
is the fact that
liquid $^3$He can be locally efficiently heated with thermal neutrons
(Fig.~\ref{f.NeutronAbsorption}).


The neutrons are produced with a paraffin moderated Am-Be source of 10 mCi
activity. They are
then incident upon the $^3$He sample container. At the minimum distance of
22 cm between source
and sample, $\nu \approx 20$ neutrons/min are absorbed by the $^3$He
liquid. For thermal
neutrons the  mean free path is 0.1 mm in liquid $^3$He and thus all
absorption reactions occur
close to the walls of cylindrical container which has radius $R=2.5$ mm.
Since the incident thermal
neutron has low momentum, the 573 keV proton and 191 keV triton fly apart
in opposite directions,
producing 70 and 10 $\mu$m long ionisation tracks, respectively. The
subsequent charge recombination
yields a heated region of the normal liquid phase which for simplicity we
here assume to be of a
spherically symmetric shape
\cite{Schiffer-Osheroff}.

The bubble of normal fluid cools by the diffusion of quasi-particle
excitations out into  the
surrounding superfluid with a diffusion constant $D \approx v_F l$ where
$v_F$ is their Fermi
velocity and $l$ the mean free path.  The difference from the surrounding
bulk temperature $T_0$
as a  function of the  radial distance $r$ from the centre of the bubble is
given by
\begin{equation} T(r,t) - T_{0} \approx {E_0\over C_v} {1\over (4 \pi D t
)^{3/2}} \exp \Biggl
({-r^2\over 4Dt} \Biggr ), \label{e.1} \end{equation}
where $E_0$ is the energy deposited by the neutron as heat and $C_v$ the
specific heat. The maximum
value  of the bubble radius $R_{b}$ with fluid in the normal phase,
$T(r)>T_c$, is
\begin{equation} R_{b} \sim  (E_0/ C_v T_c)^{1/3} (1-T_{0}/T_c)^{-1/3}~.
\label{e.2} \end{equation}
The bubble cools and shrinks away rapidly with the   characteristic  time
$\tau_Q \sim R_b^2/D
\sim 10^{-6}$ s. The rapid superfluid phase transition leads to the
formation of  a random
network of vortices as was originally formulated by Kibble \cite{Kibble}
and later refined by
Zurek \cite{Zurek}.

The fate of nucleated vorticity depends on the bias field, the velocity
$v_s$ of the superflow
outside the bubble in the bulk $^3$He-B. The superflow is created by
rotating the
container: the normal component is clamped to corotation with the container
while the superfluid
component remains stationary. This results in a relative superfluid
velocity $v_s = \Omega
R$ at the wall of the container, when rotated at an angular velocity
$\Omega$. If  the radius of
the  vortex loop exceeds the value $r_\circ(v_s) = ( \kappa/ 4\pi v_s)
\ln{r_\circ/\xi}$, where
$\kappa=\pi\hbar/m_3$ is the circulation quantum and
$\xi=\xi_0(1-T/T_c)^{-1/2}$ is the
superfluid coherence length, the loop expands. An expanding vortex ring
eventually results in a
rectilinear  vortex line which is pulled to the center of the container. If
several vortex lines
are formed they accumulate in a vortex cluster \cite{Nucleation}.

The number of the vortex lines is monitored with NMR. In the insert of
Fig.~\ref{f.OmegaDepend}
two NMR absorption records are shown, starting from the moment when the
neutron source is
placed in position. The absorption events, which lead to nucleation, are
visible as
distinct steps. The step height gives
the number of vortex lines nucleated per event, if the neutron source is
sufficiently far from
the sample and the absorption events are well separated in time. The number
of nucleated
vortex lines per unit time increases rapidly with increasing $v_s$
(Fig.~\ref{f.OmegaDepend}).
The extrapolation to zero gives the threshold value  $v_{cn}$ plotted in
Fig.~\ref{f.Omega_cn} as
a function of $T$ for different pressures. This $v_{cn}$ is smaller and has
a different
dependence on $T$ than the critical velocity $v_c$ at which a vortex is
nucleated in the absence
of neutrons \cite{Nucleation}.


Let us estimate the critical velocity $v_{cn}$ and the number of vortices
${\cal N}(v_s)$ created in one
neutron absorption event \cite{Kibble,Zurek}. The topological defects are
nucleated during the
superfluid transition to $^3$He-B when the order parameter begins to fall
from the  false
ground-state into the true degenerate ground-states which form independently
in regions which are not causally connected. The different regions grow in
size and
ultimately the order parameter fills all space but with a multitude of
defects. These are laid down during the freeze out time $\sqrt{\tau_Q
\tau_0}$~, where $\tau_0=\xi_0 /v_F$ characterizes the relaxation
time  of the order parameter in the low temperature limit. The value of the
coherence
length $\xi$ at that point,  $\xi_0 (\tau_Q / \tau_0)^{1/4}$
\cite{ZurekReview},
determines the  initial distance between the defects, which in this
experiment is of order
1 $\mu$m.

The later evolution of the vortex network leads to a gradual increase of
the intervortex
distance $\tilde \xi(t)$ \cite{Review2}, caused  almost entirely by the
interaction
of the defects with each other, rather than by fluctuations in the order
parameter field
as before. With increasing time loops with a line length $l < \tilde \xi$
are smoothed
out while
larger loops straighten and reconnect such that the network appears scale
invariant.  For a
homogeneous system the number of loops $n(l)$ per unit length and unit
volume with line lengths $l> \tilde \xi$ is given by
\cite{VachaspatiVilenkin}
\begin{equation} n(l)=C\tilde \xi^{-3/2}l^{-5/2}\;\label{e.3} \end{equation}
where $C \sim 0.4$. For our case of finite volume of the bubble, where  all
vortices form
closed loops, our numerical simulation gives the same distribution law but
with slightly different
$C \sim 0.3$. For the average straight-line dimension $\cal D$ of a loop
we find ${\cal D} =\beta (l \tilde \xi)^{1/2}$ with
$\beta\approx 1$. The loop size
distribution in terms of ${\cal D}$ is scale invariant,  $n({\cal D})\;
d{\cal D} \approx
2C ~d{\cal D} / {\cal D}^4 $, and does
not depend on the distance between vortices. The evolution of the network
leads to an increasing lower cutoff of the distribution, ${\cal D}_{\rm
min}= \alpha
\tilde
\xi(t)$, where
$\alpha$ is of order
unity. The upper cutoff is the diameter of the bubble ${\cal D}_{\rm
max}=2R_b$.

When the average radius of curvature, determined by $\tilde
\xi$,  exceeds $r_\circ (v_s)$ the vortices start to escape from the
bubble. The number of vortices
${\cal N}(v_s)$, created per one neutron, is thus the  number of loops with
$\alpha  r_\circ(v_s) < {\cal D} <
2R_b$  within the bubble volume $V_b$:
\begin{equation} {\cal N}(v_s)=V_b\int_{\alpha  r_\circ(v_s) }^{2R_b} d{\cal
D}~
n({\cal D})= {\pi C\over 9} \left[\left({{2R_b} \over {\alpha
r_\circ(v_s)}}\right)^3
-1\right] \, . \label{e.4}
\end{equation}
 The critical velocity $ v_{cn}$ is determined by the
requirement  $ {\cal N}(v_{cn})=0$, which gives
$v_{cn}  = {\kappa \alpha \over 8\pi  R_b } \; \ln  (R_b/\xi)$. Our
numerical simulation suggests $\alpha \approx 1$. Thus $v_{cn}$ is
inversely proportional to
$R_b$ and according to Eq.~(\ref{e.2}) has the temperature dependence
$v_{cn} \propto
(1-T_{0}/T_c)^{1/3}$, in agreement with the measurements in
Fig.~\ref{f.Omega_cn}.
In terms of $ v_{cn}$ one obtains the universal curve
\begin{equation} {\cal N}(v_s/v_{cn})  = {\pi C\over
9}\left[\left({v_s\over v_{cn}}\right)^3 -1\right]~.\label{e.5} \end{equation}
 Eq.~(\ref{e.5}) reproduces the observed  cubic velocity dependence and
gives the
correct order of magnitude estimate of ${\cal N}(v_s)$, as seen in
Fig.~\ref{f.OmegaDepend}.
It  also carries the same universality feature as the measured results in
Figs.~\ref{f.OmegaDepend} and \ref{f.Universality}: the ambient measuring
conditions depend
on what values are chosen for the temperature $T$, pressure $P$, and
magnetic field $H$,
but in the results for ${\cal N}(v_s)$ all this dependence is contained in
the single parameter
$v_{cn}$.

The experiment  demonstrates that defects are nucleated within bulk liquid
$^3$He during a rapid quench to the superfluid state. An external bias
flow, which exceeds a critical value, allows the defects to escape, to
expand and to reach a stable state in the rotating container so that they
can be detected one by one. With increasing superflow velocity smaller
loops,  which represent an earlier
stage in the evolution of the network, are extracted. This allows us
to probe the  morphology of the defect network at different stages of its
evolution. The measured number of extracted vortices as a function of
velocity is
consistent with an initial loop size distribution which characterizes
the random phases of the order parameter in the quench.  We expect to learn
more about this distribution from measuring the velocity dependence of
events which lead to a given number of extracted vortex lines. More
sophisticated
numerical simulation is expected to reveal how the network of large loops,
which exceed critical size, is modified  by the bias superflow in the late
stages of expansion.


This collaboration was carried out under the EU Human Capital and
Mobility Programme. 
AJG and TWBK thank the Low
Temperature Laboratory  for support and hospitality.



\begin{figure} \caption{Principle of the experiment and of vortex
nucleation during
neutron irradiation: a) Superfluid $^3$He-B contained in a cylinder, which
rotates at constant
angular velocity $\Omega$, is exposed to neutron radiation. b) In a neutron
absorption event a
proton and a triton are formed which fly apart in
opposite directions. Their
kinetic energies are used up in ionization reactions of $^3$He atoms. c)
The ionized particles
recombine generating heat in the form of quasiparticle excitations which
corresponds to
a substantial
part of the total reaction energy. A cigar-shaped hot region
of liquid is formed where
the temperature rises above $T_c$. d) The hot bubble cools rapidly  towards
the bulk
liquid temperature $T_0$. While passing through the superfluid transition
at $T_c$ a tangled network
of vortex lines is formed. Vortex loops which exceed a critical size start
to expand in the
superflow created by the rotation of the container whereas smaller loops
contract and annihilate. e)
An expanding vortex ring ultimately intersects the walls of the
container and gives rise to one
rectilinear vortex line. The vortex lines accumulate in the center of the
rotating container where
they are counted with NMR. }\label{f.NeutronAbsorption}
\end{figure}

\begin{figure} \caption{Nucleation of vortices during neutron irradiation.
{\it Insert:} NMR absorption as a function of running time $t$ at high and low
rotation velocity. The neutron source is turned on at
$t=0$. Each step corresponds to one neutron absorption event. The height of
the step
gives the number of nucleated vortex lines and is denoted by the adjacent
number. The high
velocity trace is recorded with the neutron source at a 2.8 times larger
distance from
the $^3$He-B sample. The bulk liquid temperature is $T_0 = 0.96\;T_c$.
{\it Main
frame:} The rate
$\dot N$ at which  vortex lines are created during neutron irradiation,
plotted as a function of the normalized superflow velocity $v_s/v_{cn}$.
Below the critical threshold at $v_{cn}$ the rate vanishes while above the
rate follows
the fitted cubic dependence $\dot N = \gamma \,  [(v_s/v_{cn})^3 - 1]$,
with $\gamma =
(1.37 \pm 0.03)$  vortex lines/min, shown by the solid line. The rate of
vortex
nucleation is the number of vortices per one neutron ${\cal N}(v_s)$
multiplied by the
neutron flux $\nu$:  $\dot N= \nu {\cal N}(v_s)$. The theoretical estimate
for ${\cal
N}(v_s)$ from   Eq.~(\protect\ref{e.5}) with $\nu \sim 20$ neutrons/min gives
$\gamma = \nu \pi C/9 \approx 2$, which is in  order of
magnitude agreement
with the experiment. The filled
data points are measured at a bulk liquid temperature
$T_0 = 0.96
\; T_c$ and the open points at $0.90\; T_c$. Both sets of data fall on the
same cubic dependence
although $v_{cn}$ differs by 50 \%
(Fig.~\protect\ref{f.Omega_cn}). The
cubic dependence on the superflow velocity is ultimately interrupted by the
spontaneous critical
velocity $v_c$  in Fig.~\protect\ref{f.Omega_cn}, which is the maximum
possible superflow velocity. The liquid pressure $P=18.0$ bar and the
magnetic field $H=
11.8$ mT are the same as in the insert. }\label{f.OmegaDepend}
\end{figure}

\begin{figure} \caption{Critical superflow velocities as a function of the
normalized bulk liquid
temperature $1-T_0/T_c$ and pressure $P$ of the liquid $^3$He-B sample in a
magnetic field of $H=11.8$  mT.
{\it Solid lines:}  Critical value of superflow velocity  $v_{cn} \propto
(1-T/T_c)^{1/3}$ at which
vortex creation starts in the presence of the neutron source. {\it Dashed
lines:} Critical threshold
for superflow  $v_c \propto (1-T/T_c)^{1/4}$ in the absence of the  neutron
source
\protect\cite{Nucleation}. } \label{f.Omega_cn}
\end{figure}

\begin{figure} \caption{Scale invariance of the vortex nucleation law during
neutron irradiation in different ambient conditions: in
Eq.~(\protect\ref{e.5}) the
dependence on temperature, pressure, and magnetic field is contained in the
critical
velocity $v_{cn}$. In this plot the nucleation rate $\dot N$  is shown as a
function of
$v_s^3$. The intercept with the horizontal axis gives $v_{cn}$ for each
line while the
common intercept with the vertical
axis, $\gamma = (1.46 \pm 0.12)$ vortex lines/min,  defines the numerical
prefactor in
Eq.~(\protect\ref{e.5}). The experimental variables $P,\;T$ and $H$ are
listed in the legend for each set of data points.  }\label{f.Universality}
\end{figure}

\end{document}